\documentclass{aa}
\usepackage{url}  

\usepackage{graphicx}
\usepackage[outercaption,rightcaption]{sidecap}
\usepackage[varg]{txfonts}
\usepackage{color}             
\usepackage[colorlinks=true,allcolors=blue]{hyperref} 
\newcommand{\etal}{et al.}
\newcommand{\be}{\begin{equation}}
\newcommand{\ee}{\end{equation}}
\newcommand{\bea}{\begin{eqnarray}}
\newcommand{\eea}{\end{eqnarray}}
\newcommand{\beq}{\begin{eqnarray}}
\newcommand{\eeq}{\end{eqnarray}}

\newcommand{\RSUN}{R$_{\odot}$}

\newcommand{\DG}{\ensuremath{^\circ}}

\renewcommand{\bf}{}

\graphicspath{ {./}{./images/}{./montage/}}
\begin{document}
        
\title{High-resolution observations with ARTEMIS/JLS and the NRH\thanks{This work is dedicated to the memory of Costas Caroubalos (1928-2021), founder of the ARTEMIS radiospectrograh.}}
\subtitle{IV. Imaging  spectroscopy of spike-like structures near the front of type-II bursts}
\author{        S. Armatas \inst{1},
                C. Bouratzis \inst{1},
                A. Hillaris \inst{1},
                C.E. Alissandrakis \inst{2},
                P. Preka-Papadema \inst{1}
                A. Kontogeorgos \inst{3},
                P. Tsitsipis \inst{3},
        \and      
                X. Moussas \inst{1}}
\offprints{S. Armatas}
\institute{Department of Physics, National and Kapodistrian University of Athens, 15783 Athens, Greece\\ 
\email{sarmatas@phys.uoa.gr}
        \and Department of Physics, University of Ioannina, 45110 Ioannina, Greece
        \and University of Thessaly, 35100 Lamia, Greece}
\authorrunning{Armatas et al.}
\titlerunning{Imaging of spike-like structures near the front of type-II bursts}
\date{Received .....; accepted ......}
\abstract
        {Narrowband bursts (spikes) {\bf are very small duration and bandwidth bursts which} appear on dynamic spectra from microwave to decametric frequencies. They are believed to be manifestations of small-scale energy release through magnetic reconnection.}
{We study the position of the  spike-like structures relative to the front of type-II bursts and their role in the burst emission.}
{We used high-sensitivity, low-noise dynamic spectra obtained with the acousto-optic analyzer (SAO) of the \mbox{ARTEMIS-JLS} solar radiospectrograph, in conjunction with high-time-resolution images from the Nan\c cay Radioheliograph (NRH) in order  to study spike-like bursts near the front of a type-II radio burst recorded at the west limb during the November 3, 2003 extreme solar event. The spike-like emission in the dynamic spectrum was enhanced by means of high-pass-time filtering.}
{We identified a number of spikes in the NRH images. Due to the lower temporal resolution of the NRH, multiple spikes detected in the dynamic spectrum appeared as single structures in the images. These spikes had an average size of $\approx$ 200\arcsec\ and their observed brightness temperature was 1.4 to $5.6\times10^9$\,K, providing a significant contribution to the emission of the type-II burst front. At variance with a previous study on the type-IV associated spikes, we found no systematic displacement between the spike emission and the emission between spikes. At 327.0\,MHz, the type II emission was located about 0.3~\RSUN\   above the pre-existing continuum emission, which, in turn, was located 0.1~\RSUN~above the western limb.}
{This study, combined with our previous results, indicates that the spike-like chains aligned along the type II burst MHD shock front are not a perturbation of the type II emission, as in the case of type IV spikes, but a manifestation of the type II emission itself. The preponderance of these chains, together with the lack of isolated structures or irregular clusters, points towards some  {form} of small-scale magnetic reconnection, organized along the type-II propagating front.}
\maketitle
%
\section{Introduction}\label{Intro}

Type II radio bursts are the electro-magnetic signature of magnetohydrodynamic (MHD) shock fronts propagating in the solar corona; these are driven either by CMEs, front or flank, or by flare blasts \citep{Vrsnak08,Pick08}. These MHD shocks accelerate high-energy electrons from which the type-II burst emission originates. This emission  appears  on the dynamic spectra, in the metric and higher wavelengths, as slowly drifting lanes (the backbone) from high to low frequencies \citep{Roberts59,Krueger79},  thus tracing the MHD outward propagation. Furthermore, the shock-associated suprathermal electrons generate the fine structures observed near the front of the type-II bands. These include the so-called herringbones \citep{Roberts59} and slow-drift fiber-like structures  \citep{Chernov1997, Chernov2007b}. 

\begin{SCfigure*}
\includegraphics[width=0.70\textwidth]{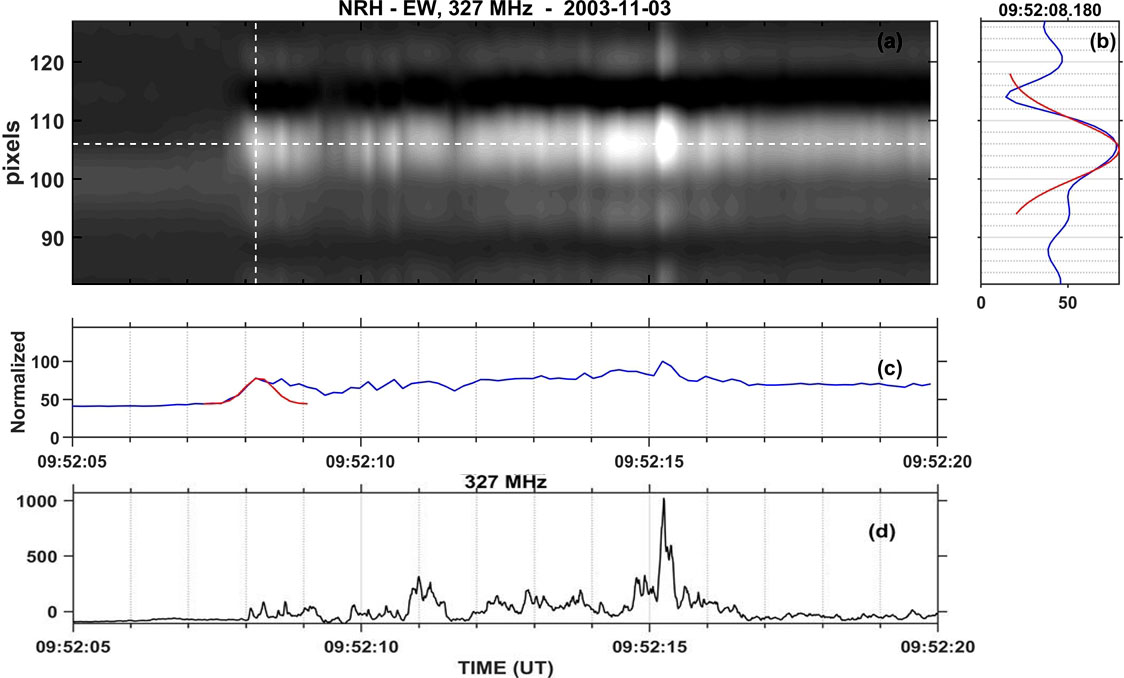}
\caption{Measurement of the duration and size  of an individual spike  in the NRH 1D recording. Panel ({\bf  b}) shows the NRH total normalized flux  in the east-west direction and panel ({\bf  c}) the temporal profile at 327  MHz. In panel ({\bf  d}) we present the Artemis-JLS/SAO time profile at 327  MHz for comparison. The red curves in panels ({\bf  b}) and ({\bf  c}) trace the gaussian fit  used in the duration and width measurement of the spikes. The bright bands above and below the main one are due to sidelobes.}
\label{NRHSpikeMeasurement}
\end{SCfigure*}
\begin{figure*}
\begin{center}
\includegraphics[width=0.90\textwidth]{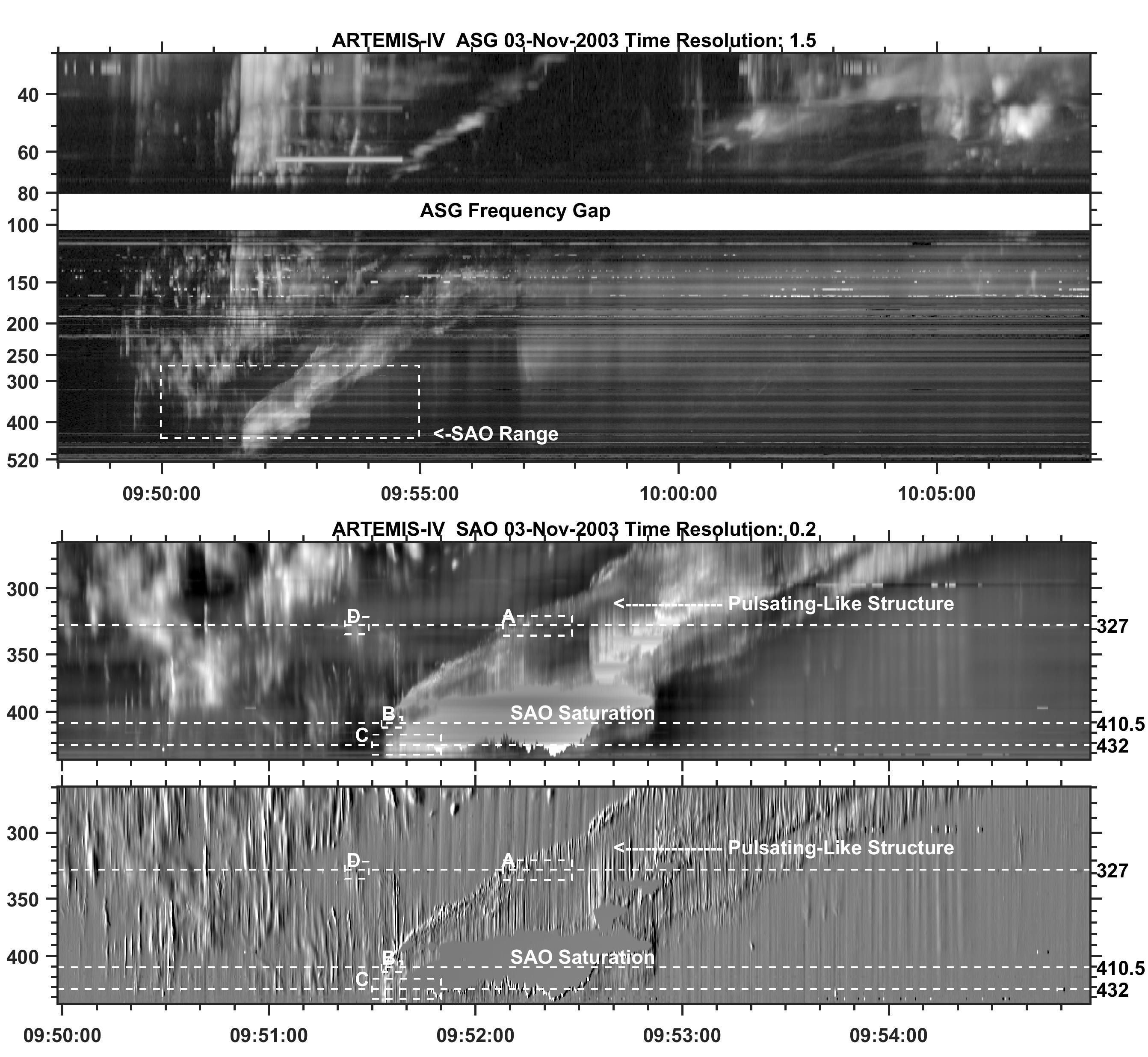}
\end{center}
\caption{Event SOL2003-11-03T09:43:20 recorded by the ARTEMIS-JLS/ASG and ARTEMIS-JLS/SAO receivers. The Dynamic spectrum of the entire event from ARTEMIS-JLS/ASG is presented in the upper panel. In the middle panel, we show the SAO dynamic spectrum  and in the lower panel, its time derivative inside the 5 min box of the upper panel. The dashed white horizontal lines on the spectrum mark three of the frequencies (327, 410.5, and 432  MHz) of the Nan\c cay radioheliograph (NRH) channels.  The  boxes indicate selected segments of spike chains, (A, B, C \& D) which are studied in this work (shown enlarged in figures \ref{SAO_Ab}, \ref{SAO_Bc}, and \ref{SAO_Ca}). }
\label{Spike_like}
\end{figure*}

A new type of type-II associated fine structure was reported in  a recent work by \citet[~hereafter Paper I]{Armatas2019}. Using the high-sensitivity (10  ms resolution) receiver of the ARTEMIS/JLS\footnote{Appareil de Routine pour le Traitement et l'Enregistrement Magnetique de l' Information Spectral/Jean-Louis Steinberg} (former ARTEMIS-IV)   radio spectrograph  \citep{Caroubalos01,Kontogeorgos06}, they detected narrowband spike-like structures, which were shorter than 100  ms and embedded in type II burst harmonic emission. An interesting result of Paper I was that most of the spikes detected were aligned in chains tracing the type II lanes with very few isolated spike-like structures observed.  The average parameters of the type-II associated spike-like structures were found to be very close to the corresponding parameters of type IV-associated spikes. A similar fine structure was also recorded by the LOFAR for the  2014 August 25 type-II radio burst \citep{Magdalenic2020} and by the IPRT/AMATERAS for the  2014 April 25 event \citep{TanBaolin2019}.

In this work, we extend the study presented in Paper I by using combined ARTEMIS-JLS and Nan\c cay Radioheliograph (NRH) to study spike-like bursts at the front of a type-II radio burst. Out of the four events studied in Paper I, there is only one, observed on November 3 2003 (\mbox{SOL2003-11-03T09:43:20}), that was recorded by the Nan\c cay Radioheliograph. The present {\bf article} is organized as follows: In Sect \ref{InstrObs}, we present the observations, data selection, and analysis. In Sect.  \ref{RS}, we give our results and we  conclude with a discussion in Sect. \ref{DisC}.

\section{Observations and data analysis} \label{InstrObs}

The basic data used in this study are high and medium time resolution dynamic spectra recorded by the ARTEMIS-JLS solar radio-spectrograph at Thermopylae \citep{Caroubalos01, Caroubalos06, Kontogeorgos}. ARTEMIS-JLS consists of a \mbox{7~m} moving parabolic dish covering the metric range and a dipole aerial adapted to the decametric range which was added in October 2002. Two receivers operate in parallel: a sweep frequency analyzer (ASG) covering the \mbox{650-20  MHz} range in 630 channels with a cadence of \mbox{10~samples/sec} and a high sensitivity multi-channel acousto-optical analyzer (SAO), which covers the \mbox{270-450  MHz} range in 128 channels with a high time resolution of \mbox{100~samples/sec}. {\bf The narrow-band, high-time-resolution SAO recordings are used in the analysis of the fine temporal and spectral structures}. As in previous works, a high pass Gaussian filter of 1\,s width in time was applied to enhance the fine structures. The broadband, medium-time-resolution data of the ASG, on the other hand, are used for the detection and analysis of radio emission from the base of the corona to $\sim2$\RSUN. We note that the ASG data were used in the overview of event SOL2003-11-03T09:43:20 in Paper I.

Images of the radio emission {\bf were} obtained from the Nan\c cay Radio Heliograph (NRH) \citep{Kerdraon97,KleinKerdraon2011}, which records two-dimensional (2D) images of the Sun at five frequencies (164, 236.6, 327, 410.5, and 432\,MHz) with a time resolution of 150 ms. All five frequencies are within the spectral range of the ASG, while the last three are also within the range of the SAO; the latter were used in conjunction with the high resolution dynamic spectra in our combined type II associated spike-like bursts analysis. 

We obtained 2D NRH images from the Nan\c cay site\footnote{https://rsdb.obs-nancay.fr/}. From the 2D images, we computed 1D images by integrating in the (east-west) EW and (north-south) NS directions; these are more convenient for comparison with dynamic spectra (see Figs. \ref{SAO_Ab}, \ref{SAO_Bc}, and \ref{SAO_Ca}) and facilitates the measurement of the duration. 

On the 150  ms NRH 1D images, we measured the duration and the size of spike-like bursts. The identification of individual bursts was done by inspection, and the width was measured after fitting the positional and spectral profiles with a Gaussian at  full width at half maximum (FWHM, see Fig. \ref{NRHSpikeMeasurement}); the method is similar to the one used by \citet[their Fig. 3]{Bouratzis2016} in their study of type-IV associated spikes.

\section{Results} \label{RS}%

In this section, we present results from the analysis of simultaneous SAO/NRH observations. We start with an overview of the 2003 November 03 event and we proceed with the study of spikes.

\subsection{Overview of the  2003 November 03 event}\label{OView}

{\bf The SOL2003-11-03T09:43:20 complex event on 2003 November 03 took place in a period of extremely high activity on the Sun between 26 October and 18 November  2003} \citep{Veselovsky2004,chertok2005,gopalswamy2006,Ishkov2006}; it was  associated with a GOES 2F/X3.9 class limb flare in AR 10488 at heliographic cooordinates 08N\DG 77W\DG \, \citep{ChertokGrechnev2005, Veronig2005,Veronig2006}. 

An overview of the radio activity of the event based on the ARTEMIS-IV/JLS ASG receiver and Wind/WAVES was presented in  \citet[]{2005Alissandrakis}.      Paper I provided more details of the type II event, which drifted at \mbox{1.6  MHz\,$\rm{s}^{-1}$ in the 500--110  MHz range}. Metric wavelength observations of the event were also analyzed by \citet{Dauphin2005, Dauphin2006, Vrsnak06, Chernov07,Aurass2013,Aurass2014}, and \citet{Chernov_2021}.

In the upper panel of Fig. \ref{Spike_like}, we present the dynamic spectrum of the type II burst with some of its accompanying activity in the range 30-530\,MHz, from recordings of ARTEMIS-JLS/ASG with 150\,ms time resolution. In the middle panel, we present details of the type II front in the harmonic within the 5 min white box of the upper panel, this time from SAO recordings with 10\,ms resolution. We note that the type II emission was strong enough to saturate the receiver in parts of the SAO spectrum. The lower panel of Fig. \ref{Spike_like} shows the time derivative of the dynamic spectrum, which enhances the fine details, both obtained by the \mbox{ARTEMIS-JLS/SAO}. 

The type II harmonic emission started at 09:51:35 UT above 500  MHz, well within the \mbox{Artemis-JLS/SAO} frequency range and the 327, 410.5, 432  MHz Nan\c cay radioheliograph channels. {\bf It exhibited rich fine structure}. Our study is focused on the spike-like burst chains of the harmonic, as  the fundamental was outside of the \mbox{Artemis-JLS/SAO} frequency limits. In the same figure, we marked four regions ({A, B, C and D}) with spikes that are presented in detail in Sect. \ref{FS}. We note that in addition to spikes, we detected pulsation-like structures just above the saturation region of the dynamic spectrum, covering a good part of the harmonic emission (see Fig. \ref{Spike_like}).

\begin{SCfigure*}[]
\includegraphics[trim=0.0cm 0.0cm  0.0cm 0.0cm,clip,width=0.75\textwidth]{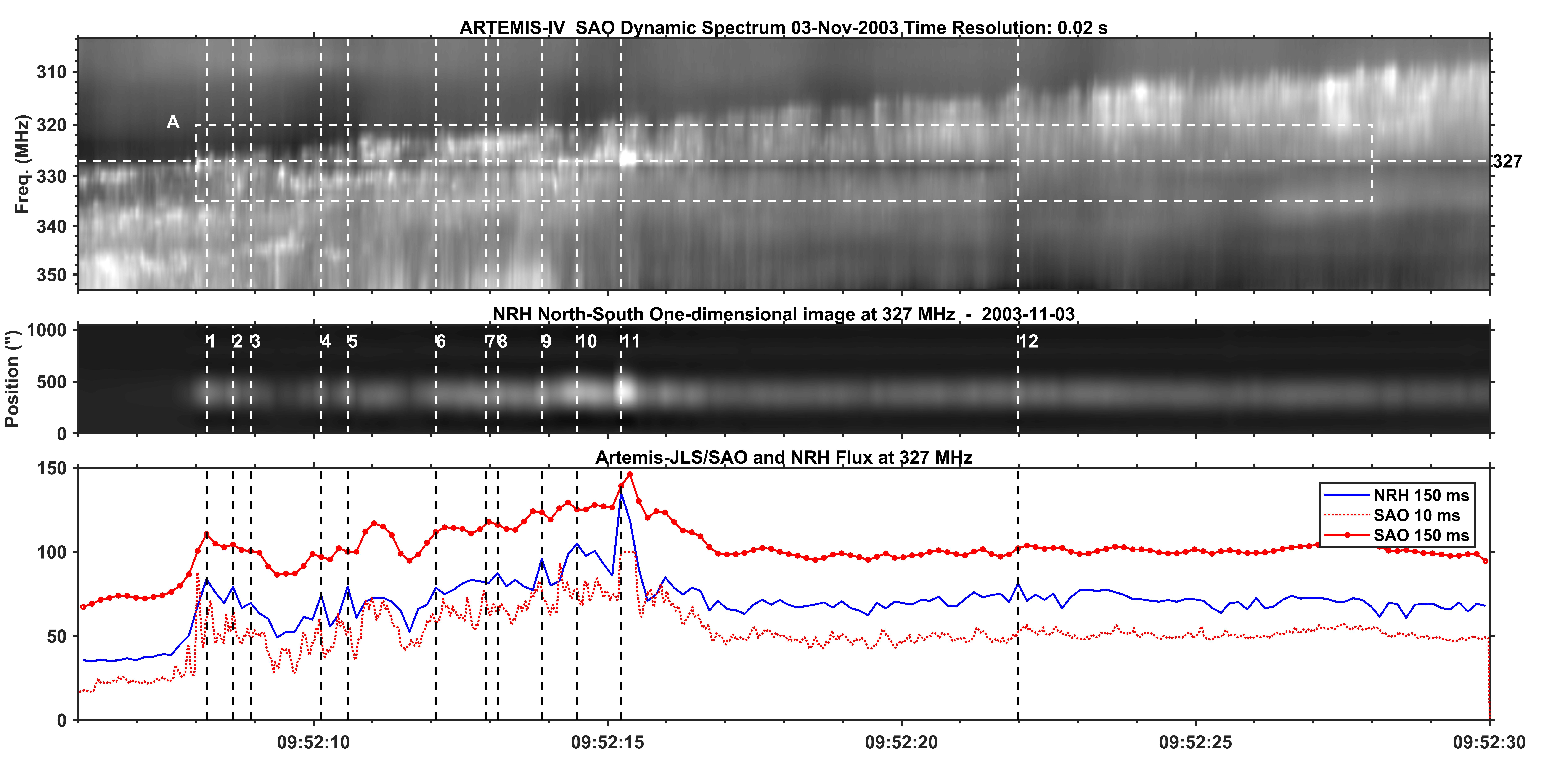}
\caption{{\bf Spike chain A recordings from SAO and NRH.} Upper panel: SAO dynamic spectrum for spike chain A (box in Fig. \ref{Spike_like}). The corresponding frequency of the NRH is marked on the right.  The next panel  shows the NS 1D NRH images at 327  MHz. In the last panel, we show NRH (blue) and SAO (red) normalized total flux time profiles at 327  MHz and at 150 ms resolution are presented; a second SAO high-resolution (10 ms) profile (red dotted) is included for comparison. Twelve spike-like structures, numbered (1-12), are marked by vertical dashed lines.}
\label{SAO_Ab}
\end{SCfigure*}
\begin{SCfigure*}[]
\includegraphics[trim=0.0cm 0.0cm  0.0cm 0.0cm,clip,width=0.75\textwidth]{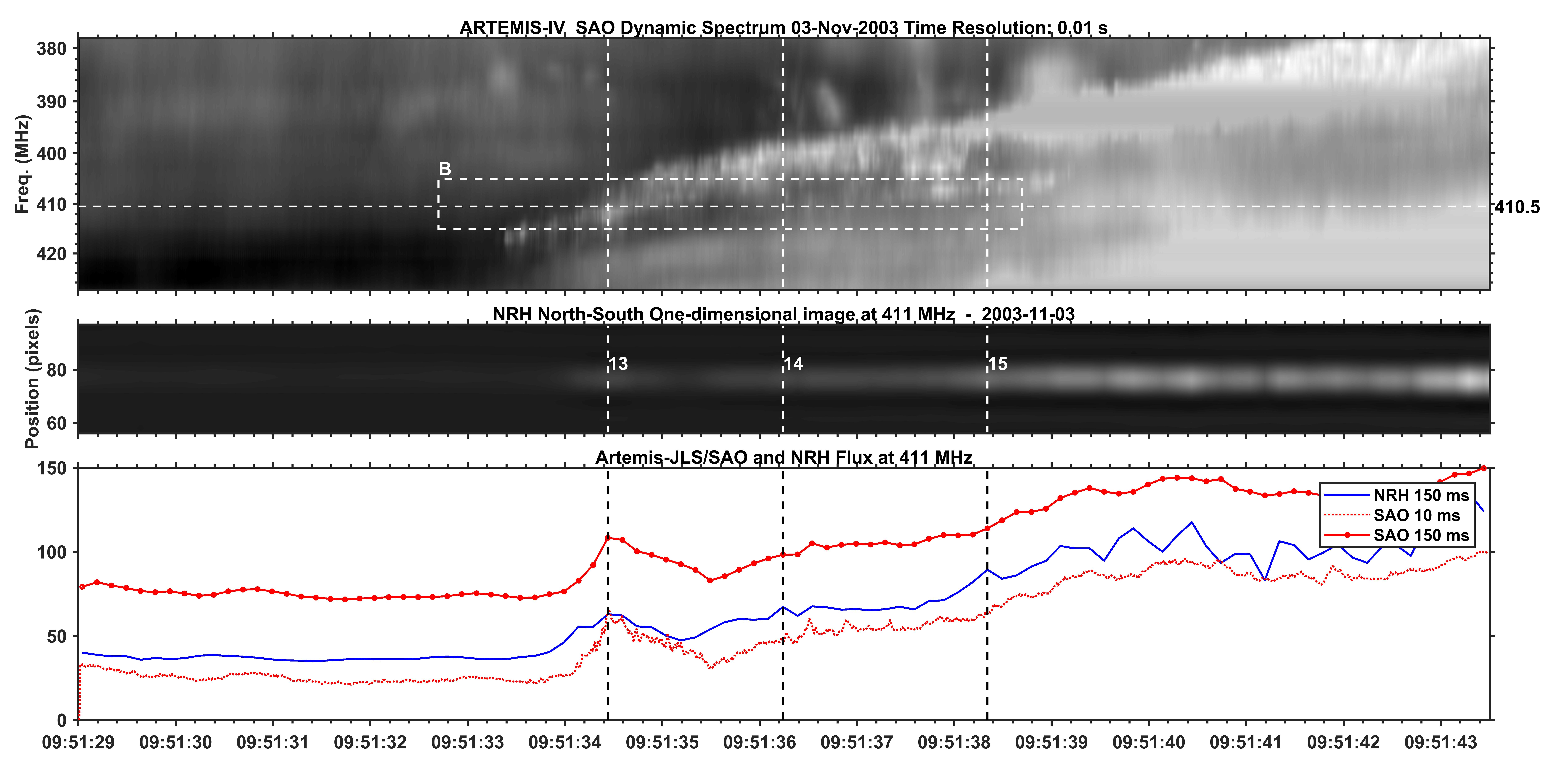}
\caption{{\bf Spike chain B recordings from SAO and NRH. Same as in Fig.} \ref{SAO_Ab} for the Spike Chain B at 410.5  MHz. Three spike-like structures, numbered (13-15), are marked by vertical dashed lines.}
\label{SAO_Bc} 
\end{SCfigure*}
\begin{SCfigure*}[]
\includegraphics[trim=0.0cm 0.0cm  0.0cm 0.0cm,clip,width=0.75\textwidth]{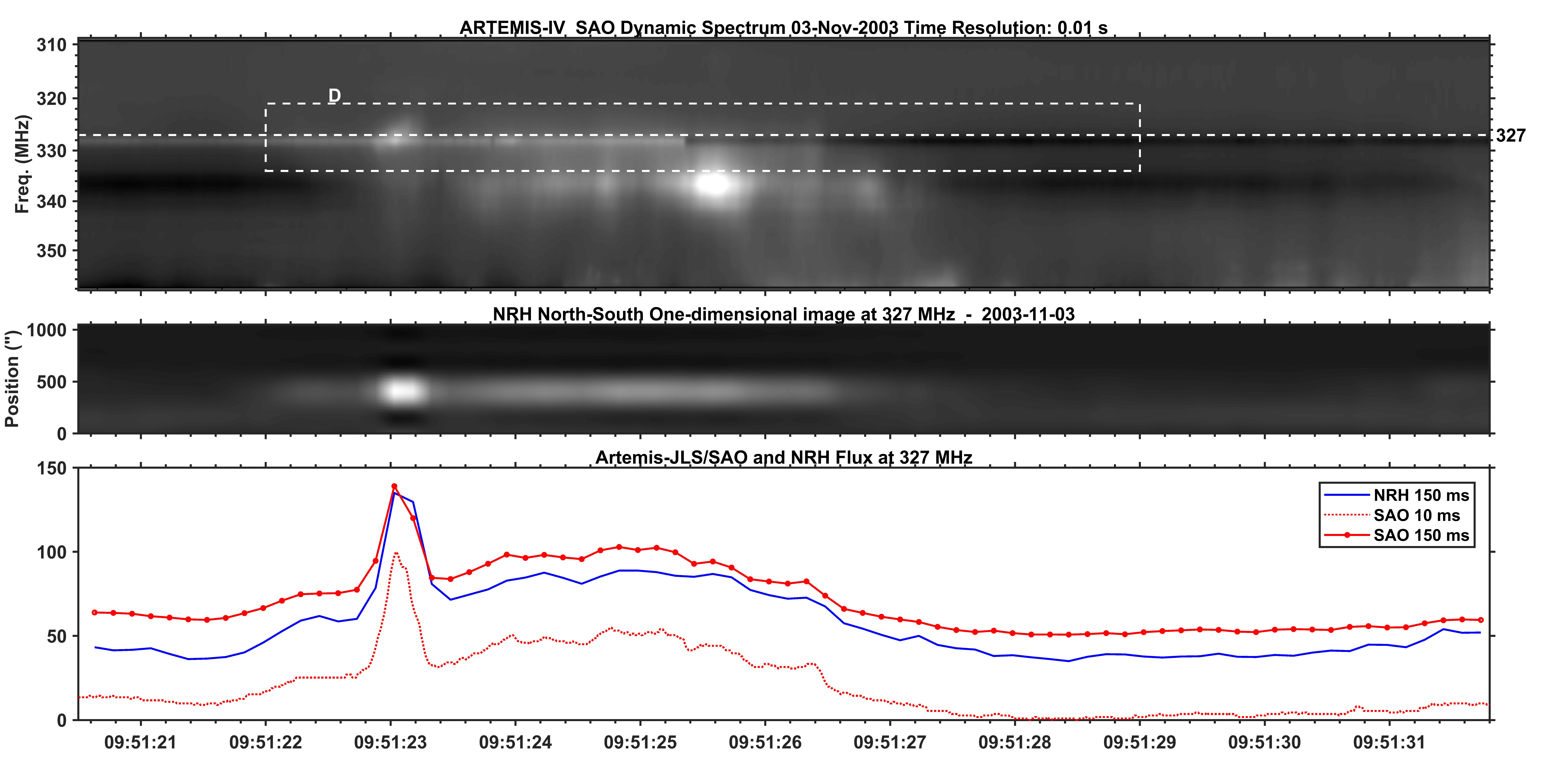}
\caption{{\bf Spike chain D recordings from SAO and NRH. Same as in Fig.} \ref{SAO_Ab} for the Spike Chain D at 327\,MHz.}
\label{SAO_Ca} 
\end{SCfigure*}

\begin{figure*}
\begin{center}
\includegraphics[width=\textwidth]{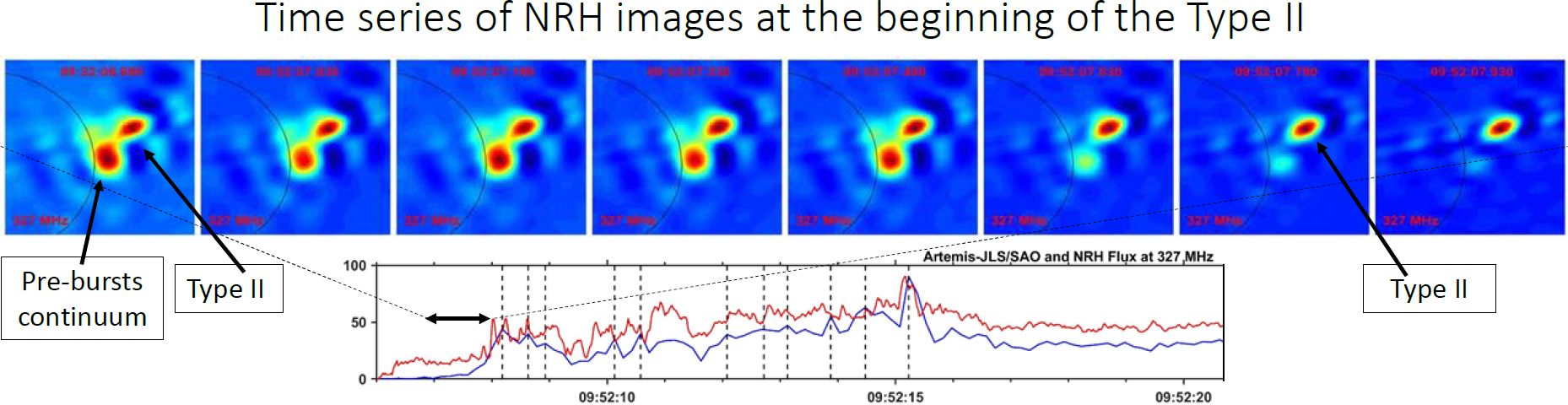}
\end{center}
\caption{{\bf NRH \& SAO images at the beginning of the type II emission at 327.0\,MHz.} Top: Sequence of NRH 2D images near the start of the Type II at 327\,Mz. The color table of each image is normalized to the corresponding minimum and maximum intensity. The black arch marks the photosheric limb. Bottom: Flux as a function of time for NRH (blue) and for SAO (red). The time range of the images is marked by the double arrow. }
\label{327Siart}
\end{figure*}

\begin{figure}[h]
\begin{center}
\includegraphics[width=.475\textwidth]{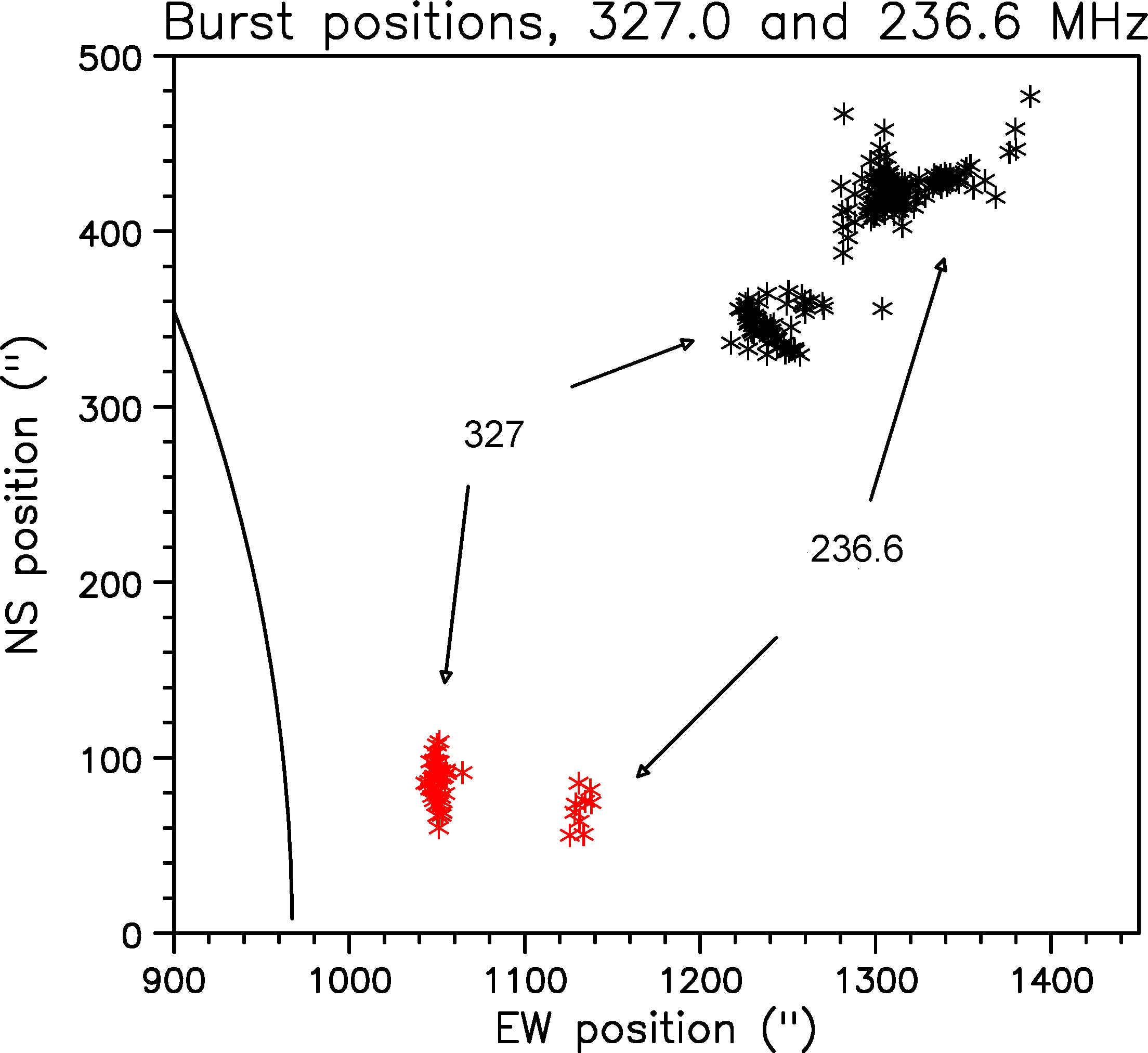}
\end{center}
\caption{Source positions from 09:52:00 to 09:52:16 UT at 327.0\,MHz and from 09:52:27 to 09:53:00 UT at 236.6\,MHz, before (red) and during (black) the type II. The black arch marks the photospheric limb.}
\label{SourcePos}
\end{figure}

\begin{SCfigure*}
\includegraphics[trim=0.0cm 0.0cm  -0.25cm 0.0cm,clip,width=.70\textwidth]{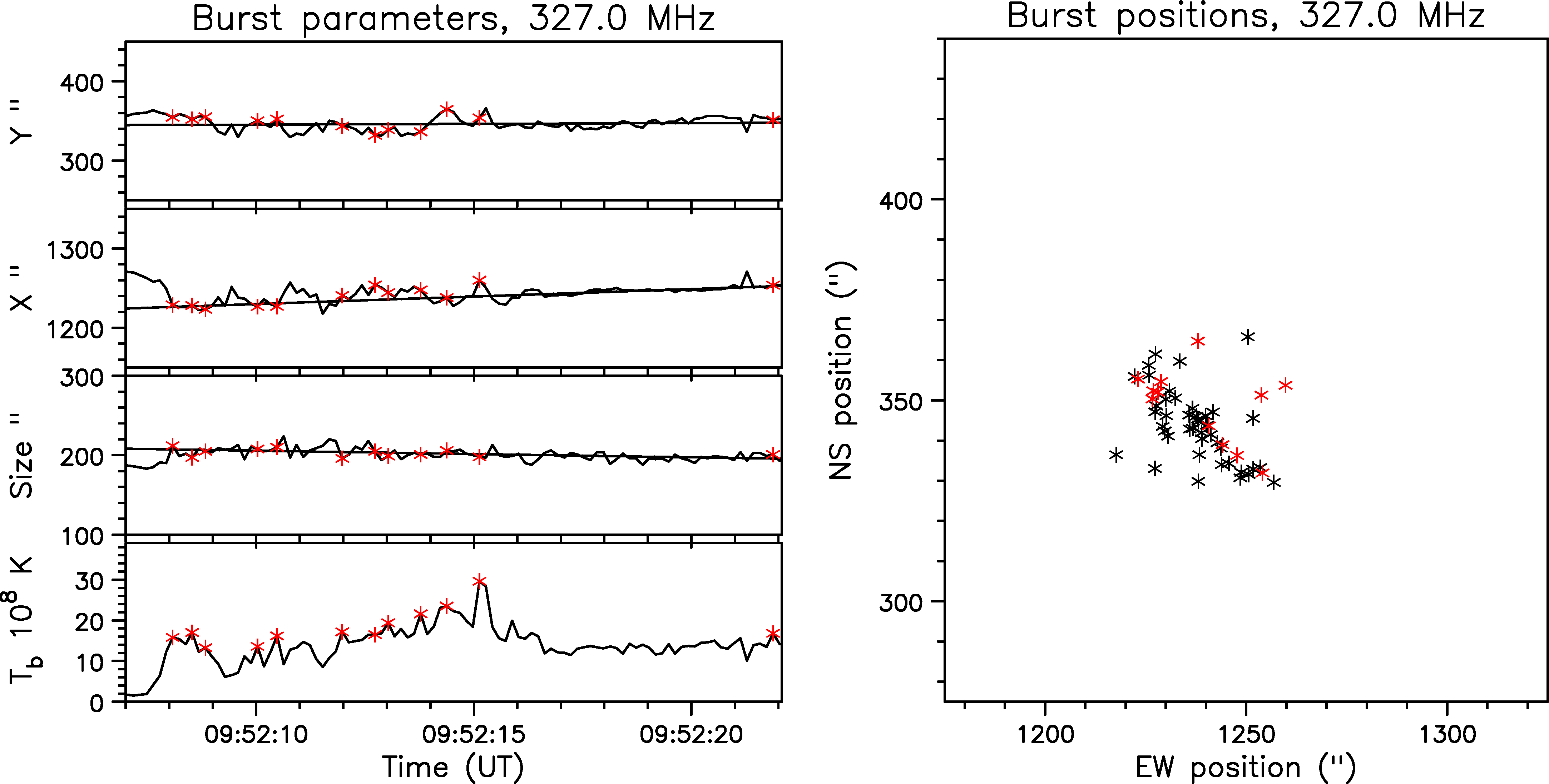}
\caption{ Brightness temperature, size, and position of sources as a function of time at 327.0\,MHz (left). Size is the geometric average of the axes of the fitted ellipse. Straight lines are the result of linear regression. Positions of the sources on the plane of the sky after 09:52:08 UT (right). Spike sources are marked as red asterisks in both plots.}
\label{SpikePos}
\end{SCfigure*}

\begin{figure}[h!]
\begin{center}
\includegraphics[trim=0.0cm 0.0cm  0.0cm 0.05cm,clip,width=.50\textwidth]{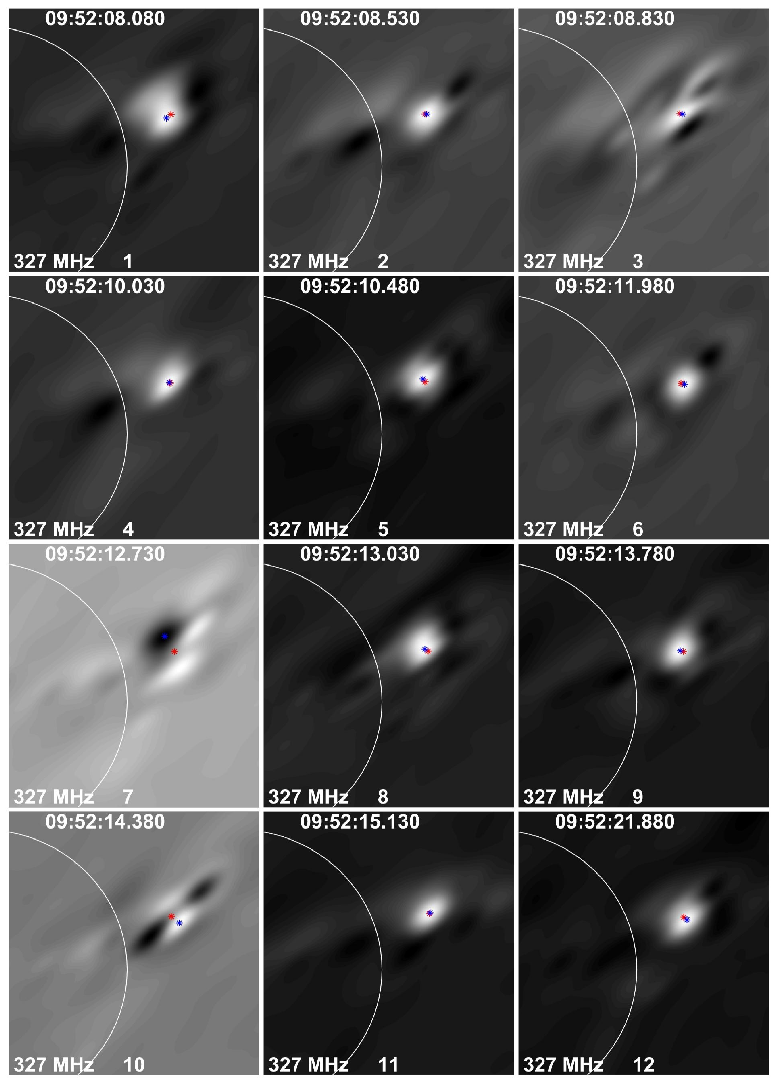}
\end{center}
\caption{Running difference images of the 12 identified spikes at 327\,MHz. The red asterisks mark the position of the maximum of the radio source, the blue asterisks that of the running difference. The white arch corresponds to the solar limb.}     
\label{Spikeimages327}
\end{figure}

\begin{figure}
\begin{center}
\includegraphics[trim=0.0cm 0.1cm  0.0cm 0.05cm,clip,width=.50\textwidth]{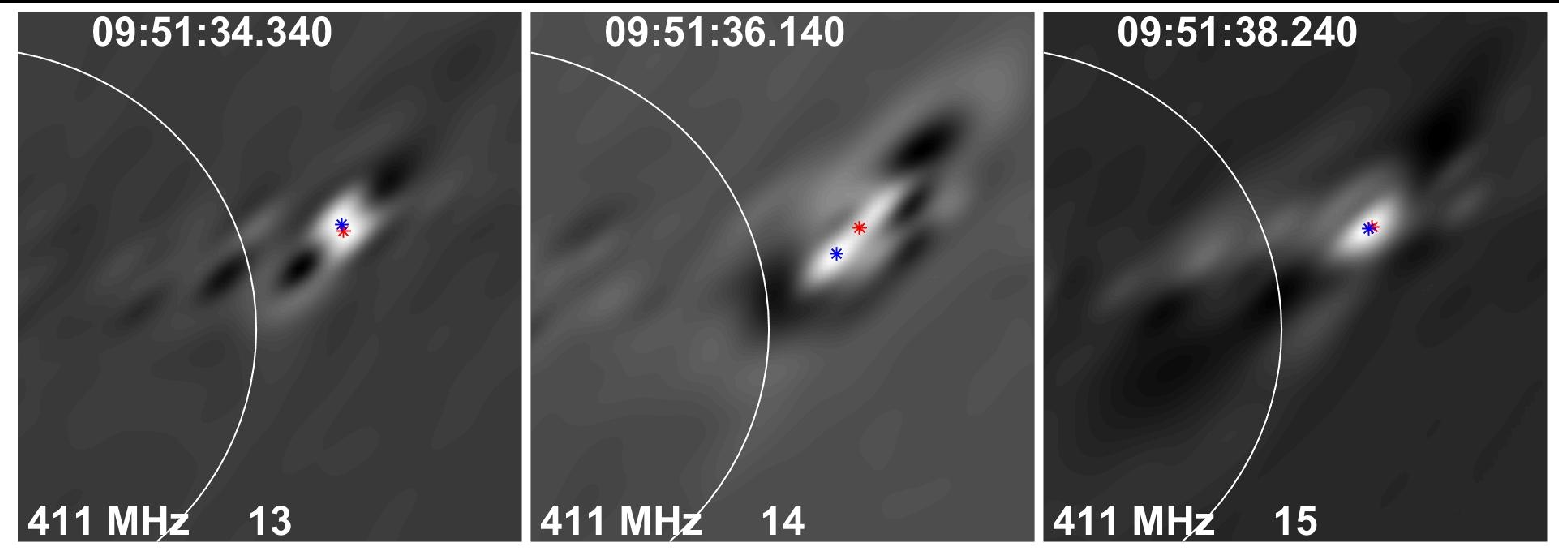}
\end{center}
\caption{{\bf Running difference images of the 3 identified spikes at 410.5 MHz. Same annotation as in Fig. \ref{Spikeimages327}}}
\label{Spikeimages410.5}
\end{figure}

\subsection{Temporal fine structure in dynamic spectra and images}\label{FS}
Due to the frequency drift of the type II, a relatively small fraction of the spikes detected in the SAO spectra are visible in the fixed frequency NRH images. The detection of the spikes {was done by inspection of the peaks in the NRH flux and 1D images and comparison with the SAO time profiles.} The process is illustrated in the Figs. \ref{SAO_Ab} and \ref{SAO_Bc}, where vertical dashed lines indicate the detected spike-like structures in the flux profiles, the 1D images, and the dynamic spectra. 

Twelve spike-like bursts which are part of two chains that trace the type II lane front at 327 MHz were identified in  spectral region A of Fig. \ref{Spike_like}. They are numbered 1-12 in Fig. \ref{SAO_Ab} which, in addition to the dynamic spectrum, shows the NS 1D NRH intensity as a function of position and time. The bottom panel of the figure shows plots of normalized flux as a function of time for both instruments.  

Near the type II front at 410.5\,MHz (region B), three spike-like bursts were detected and marked 13, 14, and 15 in Fig. \ref{SAO_Bc}. The structures beyond spike number 15 on Fig. \ref{SAO_Bc} were excluded as they were part of the pulsation-like structures just above the SAO saturation area (see Fig. \ref{Spike_like}). The dynamic spectrum at 432\,MHz (region C)  reached saturation just after the appearance of the type II front, thus we could see no spikes there. The burst in spectral region D of  Fig. \ref{Spike_like}, which preceded the harmonic band of the type II at 327\,MHz, resemble extremely narrowband type IIIs rather than spikes; nonetheless, we included it in our study for the purposes of comparison with the spike-like structures (Fig. \ref{SAO_Ca}).

We note at this point that the time resolution of the  \mbox{Artemis-JLS/SAO} is more than a factor of ten higher than that of the NRH; therefore, certain features that were detected and measured as single bursts in the NRH 1D images correspond, in fact, to two or more bursts in the dynamic spectrum. An example is shown in  the lower panels of Figs. \ref{SAO_Ab}, \ref{SAO_Bc}, and \ref{SAO_Ca} where the time profiles, at 327 and 410.5  MHz, of the NRH and the SAO at 150 ms resolution are plotted, with an SAO high-resolution (10 ms) profile for comparison. In addition, s{everal} NRH peaks correspond to multiple high resolution SAO peaks. 

Due to the difference in time resolution, the measured duration  from the 1D NRH images  is expected to be greater than the average duration of spikes measured by the SAO. Thus, the duration of the spike-like structures was found to be in the range of 150-600\,ms, while in Paper I, the average duration was reported to be less than 100 ms. This is due to the fact that individual spikes recorded by the ARTEMIS/JLS SAO receiver cannot be clearly identified in the 1D plot of the Nan\c cay Radioheliograph, as discussed above. 

\subsection{Burst positions and structure}
NRH 2D images give important information on the position of the emitting sources. Figure~\ref{327Siart} shows images at the very beginning of the type II emission at 327.0\,MHz. We note that as the type II emission increases, the relative intensity of the pre-existing continuum source diminishes; moreover, the type II source is higher than the continuum source, displaced by about 5.3\arcmin\ or 0.3\RSUN, whereas their position angles differ by 13\degr. The projected location of the pre-type II emission was  0.1\RSUN~above the west limb.

For a more detailed measurement of source positions and sizes, we performed a least-squares fit with a Gaussian model. The positions derived in this way from the 327.0\,MHz images are plotted in Fig.~\ref{SourcePos}, where, again, the shift between the type II and the pre-existing emission is clearly visible.

\begin{table}[h]
\begin{center}
\caption{Parameters of spikes measured from 2D NRH images.}
\label{2003c}
\begin{tabular}{lrrr}
\hline
Frequency (MHz)                           &                      327                                              &               410.5                                                           &                 327      \\
                                Region                                                     &                     A                                                  &             B                                                                       &                 D             \\
                        \hline                                                                                            
                                Source position, (\arcsec)        &                                                                              &                                                                            &                                       \\
                                EW                                                      &        1240$\pm${12}    &  1240$\pm${~3}               &                1225               \\
                                NS                                                       &       350$\pm${~~9}      &  330$\pm${~1}         &            380                       \\     
                        \hline           
                                Source size, (\arcsec)           &                                                                           &                                                                              &                                               \\
                                B$_{maj}$                                                 &                      235$\pm${9}                      &      205$\pm${2}                                        &     285                        \\
                                B$_{min}$                                                 &                      175$\pm${8}              &      150$\pm${4}                                       &         200                            \\
                                Position Angle ($^{\circ}$)              &                       -58$\pm${4}                               &     -55$\pm${3}                                        &     -58                              \\
                                \hline       
                                $T_{b}$, (10$^9$ K)               &                                                                              &                                                                          &                                             \\
                            Range                                                          &              1.4 - 4.5                                     &    3.3 - 5.6                                           &         1.3                      \\
                
                        \hline
                \end{tabular}
        \end{center}
\end{table}

 \begin{table}[h]
         \begin{center}
                \caption{NRH beam size.}
                \label{table_beamsize_NRH}
                \begin{tabular}{lrrrrr}
                        \hline
                                Frequency (MHz)                 & 164   & 236                     &                       327                     &                         410.5    &              432     \\
                                Region                                            &                      &                                 &                     A                            &                   B                  &    {C}      \\
                        \hline
                                Major Axis                      & 480\arcsec    & 329\arcsec       &  236\arcsec                  &    193\arcsec                 & 171\arcsec                                 \\
                                Minor Axis                      & 205\arcsec    & 145\arcsec      &               99\arcsec               &        85\arcsec                               & 78\arcsec                           \\
                        \hline                                                                                            
                                Position Angle          & -55{$^\circ$} &       -55{$^\circ$}   &                 -55{$^\circ$}           &               -55{$^\circ$}   &         -55{$^\circ$}            \\
                                
                        \hline
                \end{tabular}
         \end{center}
                
\end{table}

The source positions at 236.6\,MHz are also plotted in Fig.~\ref{SourcePos}. Although this frequency is outside the SAO range, the ASG spectrum (Fig.~\ref{Spike_like}, top) shows clearly the type II front; moreover, the NRH images show a lot of short timescale structures, some of which might be spikes.
As in the case of 327.0\,MHz, the type II source was located higher than the pre-existing continuum, this time displaced by about 6.4\arcmin, or 0.4~\RSUN~and their position angles are almost identical. Overall, both the continuum and the Type II sources at 236.6\,MHz were above the corresponding 327.0\,MHz structures, as expected; the {type-II} positions are consistent with \citet[~their Fig. 7]{Dauphin2006}.  

An important question is whether the spike emission is at the same location as the emission between spikes or, as in the case with Type IV associated spikes, it is located at the periphery of the background emission \citep{Bouratzis2016}. To investigate this issue, we plotted in Fig.~\ref{SpikePos} the source parameters, deduced from Gaussian fit, as a function of time (left) and their positions on the sky plane (right), marking the spikes by red asterisks. We note that after the initial intensity rise (09:52:08 UT), most sources  fall on a straight line, which reflects a systematic displacement to the west ($X$ plot in Fig.~\ref{SpikePos}, left), consistent with the upward motion of the type II front. Spikes 1-9 also fall along this line, whereas spikes 10-12 do not. A careful inspection of the upper panel of Fig.~\ref{SAO_Ab} reveals that the latter are part of a different spike chain, apparently associated to a different part of the shock front than the one which produced spikes 1-9.

Images of the 15 identified SAO/NRH spikes are given in Fig.~\ref{Spikeimages327} for 327.0\,MHz and Fig.~\ref{Spikeimages410.5} for 410.5\,MHz. For better visibility, and also  to show position differences between the spike source and the preceding emission, we used running difference images and marked the peak position of the source in the original and in the subtracted image. A preliminary remark is that some spikes are double (\# 1, 10, 14), with the two sources close to each other -- these are probably due to separate spikes at slightly different positions that are not resolved temporally by the NRH. In conformity with our previous conclusion, there appears to be no large shift between the spike and the pre-spike emission, with the exception of spikes 7 and 10.
         
The parameters derived from the Gaussian fit are given in Table~\ref{2003c}. The source size is of the order of 150\arcsec\ to 240\arcsec; thus the sources are not fully resolved, as their axes are not much larger than the NRH beam size (given in Table~\ref{table_beamsize_NRH} for all frequencies). This can be seen in their elongated shape in Figs. \ref{Spikeimages327} and \ref{Spikeimages410.5}. Their brightness temperature is on the order of 10$^9$\,K, which is a lower limit both due to the beam size and refraction effects.        

\section{Discussion and conclusions} \label{DisC}
        
We examined the characteristics of type II spike-like bursts detected from high temporal resolution recordings of ARTEMIS-JLS/SAO combined with imaging from the  Nan\c cay Radioheliograph. Complementing our Paper I results, we estimated the position, size, duration, and shift of the structures from NRH.
        
The small number of fifteen spike-like structures in the type II front that were identified in both instruments is due to the frequency drift of the type II bursts, which shifts the spike-like sources outside of the NRH channel; therefore, relatively short segments of the spike chains detected by the SAO were recorded by the NRH. 
        
We found that the durations measured from NRH data with a sampling interval of 150 ms exceeded the corresponding values from the ARTEMIS-JLS/SAO data. This indicates that most single spike burst measurements of the NRH recordings correspond to groups of spikes detected in the SAO dynamic spectrum. The size of spikes was in the 150\arcsec-240\arcsec\ range, not fully resolved in the NRH images, whereas their observed brightness temperature was in the  1.4 to $5.6\times10^9$\,K range.

Our detailed investigation of source positions at 327.0\,MHz showed that the type II emission was located $\sim0.3$\,\RSUN~above the pre-existing continuum source; similar results were obtained for 236.6\,MHz. Most type II emission peaks were located along a straight line, indicating upward motion of the source. The same trend was observed for all the spike sources, with the exception of three, which were part of a different spike chain. 

We found no substantial position difference between spike and inter-spike emission, thus type II spikes appear to be associated with the entire emitting region. This is in contrast to the type IV spikes that are located at the periphery of the background source \citep{Bouratzis2016} and, hence, are a perturbation on top of the continuum emission. This indicates that {spikes constitute the principal} radio signature of the MHD shock \mbox{type-II} radio emission (see Paper I), with {their} chains {delineating} the shock front emission; this front appears continuous on low-resolution dynamic spectra.  {Still}, the type-II and type-IV associated spikes are quite similar {(see Table 1 in Paper I), possibly hinting towards a similar emission mechanism. 

We found evidence that different spike chains make up  parallel drifting lanes. These combined, probably correspond to the  ``band-split of the band-split'' image of the type-II burst front, reported by \citet{Magdalenic2020}, since the continuity of the front line depends on the dynamic spectrum resolution, as mentioned in the previous paragraph. A possible interpretation of the multiple lanes responsible for the band split is provided in \citet{Chernov_2021}, asserting that it may be due to several sources formed at different positions of the large-scale type-II shock front.

One of the prime candidates for the interpretation of spikes is the emission associated to plasma instabilities due to electron beams originating in a sequence of small-scale reconnections associated to different regions of the shock front. The preponderance of spike-like burst chains that are almost parallel to one another and the scarcity of isolated bursts and irregular clusters of spikes at the fronts of the type-II bursts suggest some type of organized structure, as opposed to random turbulence driven reconnection. Thus, triggering along the front of the MHD shock disturbance might be at the origin of such processes.

\begin{acknowledgements} 
The authors wish to thank  the radio monitoring service at LESIA (Observatoire de Paris) for providing data used for this study, in particular our colleagues from the Observatoire de Paris-Meudon and the NRH staff. Data from GOES, NOAA and SOHO-LASCO were obtained from the respective data bases; we are grateful to all those who contributed to the operation of these instruments and made the data available to the community. We also wish to thank  the Onassis Foundation for financial support (Grant 15153) for the continued operation of the ARTEMIS-JLS radio spectrograph and the University of Athens Research Committee for Grant 15018. 
\end{acknowledgements}


\begin{thebibliography}{31}
\expandafter\ifx\csname natexlab\endcsname\relax\def\natexlab#1{#1}\fi

\bibitem[{{Alissandrakis} {et~al.}(2005){Alissandrakis}, {Nindos}, {Hilaris},
  {Caroubalos}, \& {Artemis Team}}]{2005Alissandrakis}
{Alissandrakis}, C.~E., {Nindos}, A., {Hilaris}, A., {Caroubalos}, C., \&
  {Artemis Team}. 2005, in ESA Special Publication, Vol.~11, The Dynamic Sun:
  Challenges for Theory and Observations, ed. D.~{Danesy}, S.~{Poedts}, A.~{de
  Groof}, \& J.~{Andries},
  \href{https://ui.adsabs.harvard.edu/abs/2005ESASP.600E.106A}{106.1}

\bibitem[{{Armatas} {et~al.}(2019){Armatas}, {Bouratzis}, {Hillaris}, {Alissand
  rakis}, {Preka-Papadema}, {Moussas}, {Mitsakou}, {Tsitsipis}, \&
  {Kontogeorgos}}]{Armatas2019}
{Armatas}, S., {Bouratzis}, C., {Hillaris}, A., {et~al.} 2019,
  \href{http://dx.doi.org/10.1051/0004-6361/201834982}{\color{magenta}\aap},
  \href{https://ui.adsabs.harvard.edu/abs/2019A&A...624A..76A}{624, A76}

\bibitem[{{Aura{\ss}}(2014)}]{Aurass2014}
{Aura{\ss}}, H. 2014,
  \href{http://dx.doi.org/10.1007/s11207-014-0604-9}{\color{magenta}\solphys},
  \href{https://ui.adsabs.harvard.edu/abs/2014SoPh..289.4517A}{289, 4517}

\bibitem[{{Aura{\ss}} {et~al.}(2013){Aura{\ss}}, {Holman}, {Braune}, {Mann}, \&
  {Zlobec}}]{Aurass2013}
{Aura{\ss}}, H., {Holman}, G., {Braune}, S., {Mann}, G., \& {Zlobec}, P. 2013,
  \href{http://dx.doi.org/10.1051/0004-6361/201321111}{\color{magenta}\aap},
  \href{http://adsabs.harvard.edu/abs/2013A%26A...555A..40A}{555, A40}

\bibitem[{{Bouratzis} {et~al.}(2016){Bouratzis}, {Hillaris}, {Alissandrakis},
  {Preka-Papadema}, {Moussas}, {Caroubalos}, {Tsitsipis}, \&
  {Kontogeorgos}}]{Bouratzis2016}
{Bouratzis}, C., {Hillaris}, A., {Alissandrakis}, C.~E., {et~al.} 2016,
  \href{http://dx.doi.org/10.1051/0004-6361/201527229}{\color{magenta}\aap},
  \href{http://adsabs.harvard.edu/abs/2016A%26A...586A..29B}{586, A29}

\bibitem[{{Caroubalos} {et~al.}(2006){Caroubalos}, {Alissandrakis}, {Hillaris},
  {Preka-Papadema}, {Polygiannakis}, {Moussas}, \& \etal}]{Caroubalos06}
{Caroubalos}, C., {Alissandrakis}, C.~E., {Hillaris}, A., {et~al.} 2006, in
  American Institute of Physics Conference Series, Vol. 848, Recent Advances in
  Astronomy and Astrophysics, ed. N.~{Solomos},
  \href{http://adsabs.harvard.edu/abs/2006AIPC..848..864C}{864--873}

\bibitem[{{Caroubalos} {et~al.}(2001){Caroubalos}, {Maroulis}, {Patavalis},
  {Bougeret}, {Dumas}, {Perche}, \& \etal}]{Caroubalos01}
{Caroubalos}, C., {Maroulis}, D., {Patavalis}, N., {et~al.} 2001, Experimental
  Astronomy, \href{http://adsabs.harvard.edu/abs/2001ExA....11...23C}{11, 23}

\bibitem[{{Chernov}(1997)}]{Chernov1997}
{Chernov}, G.~P. 1997, Astronomy Letters,
  \href{http://adsabs.harvard.edu/abs/1997AstL...23..827C}{23, 827}

\bibitem[{{Chernov} \& {Fomichev}(2021)}]{Chernov_2021}
{Chernov}, G.~P. \& {Fomichev}, V.~V. 2021,
  \href{http://dx.doi.org/10.3847/1538-4357/ac1f32}{\color{magenta}\apj}, 922,
  922

\bibitem[{{Chernov} {et~al.}(2007{\natexlab{a}}){Chernov}, {Kaiser},
  {Bougeret}, {Fomichev}, \& {Gorgutsa}}]{Chernov07}
{Chernov}, G.~P., {Kaiser}, M.~L., {Bougeret}, J.-L., {Fomichev}, V.~V., \&
  {Gorgutsa}, R.~V. 2007{\natexlab{a}},
  \href{http://dx.doi.org/10.1007/s11207-007-0258-y}{\color{magenta}\solphys},
  \href{http://adsabs.harvard.edu/abs/2007SoPh..241..145C}{241, 145}

\bibitem[{{Chernov} {et~al.}(2007{\natexlab{b}}){Chernov}, {Stanislavsky},
  {Konovalenko}, {Abranin}, {Dorovsky}, \& {Rucker}}]{Chernov2007b}
{Chernov}, G.~P., {Stanislavsky}, A.~A., {Konovalenko}, A.~A., {et~al.}
  2007{\natexlab{b}},
  \href{http://dx.doi.org/10.1134/S1063773707030061}{\color{magenta}Astronomy
  Letters}, \href{http://adsabs.harvard.edu/abs/2007AstL...33..192C}{33, 192}

\bibitem[{Chertok {et~al.}(2005)Chertok, Fomichev, Gnezdilov, Gorgutsa,
  Markeev, \& Sobolev}]{chertok2005}
Chertok, I., Fomichev, V., Gnezdilov, A., {et~al.} 2005, Astronomical \&
  Astrophysical Transactions, 24, 24

\bibitem[{{Chertok} \& {Grechnev}(2005)}]{ChertokGrechnev2005}
{Chertok}, I.~M. \& {Grechnev}, V.~V. 2005,
  \href{http://dx.doi.org/10.1134/1.1862362}{\color{magenta}Astronomy Reports},
  \href{http://adsabs.harvard.edu/abs/2005ARep...49..155C}{49, 155}

\bibitem[{{Dauphin} {et~al.}(2006){Dauphin}, {Vilmer}, \&
  {Krucker}}]{Dauphin2006}
{Dauphin}, C., {Vilmer}, N., \& {Krucker}, S. 2006,
  \href{http://dx.doi.org/10.1051/0004-6361:20054535}{\color{magenta}\aap},
  \href{http://adsabs.harvard.edu/abs/2006A%26A...455..339D}{455, 339}

\bibitem[{{Dauphin} {et~al.}(2005){Dauphin}, {Vilmer}, {L{\"u}thi}, {Trottet},
  {Krucker}, \& {Magun}}]{Dauphin2005}
{Dauphin}, C., {Vilmer}, N., {L{\"u}thi}, T., {et~al.} 2005,
  \href{http://dx.doi.org/10.1016/j.asr.2005.04.092}{\color{magenta}Advances in
  Space Research}, \href{http://adsabs.harvard.edu/abs/2005AdSpR..35.1805D}{35,
  1805}

\bibitem[{Gopalswamy(2006)}]{gopalswamy2006}
Gopalswamy, N. 2006, Solar Extreme Events: Fundamental Science and Applied
  Aspects. Yerevan: Alikhanyan Physics Institute, 24, 24

\bibitem[{{Ishkov}(2006)}]{Ishkov2006}
{Ishkov}, V.~N. 2006,
  \href{http://dx.doi.org/10.1134/S0038094606020055}{\color{magenta}Solar
  System Research},
  \href{http://adsabs.harvard.edu/abs/2006SoSyR..40..117I}{40, 117}

\bibitem[{{Kerdraon} \& {Delouis}(1997)}]{Kerdraon97}
{Kerdraon}, A. \& {Delouis}, J.-M. 1997, in Lecture Notes in Physics, Vol. 483,
  Coronal Physics from Radio and Space Observations, Berlin Springer Verlag,
  ed. G.~{Trottet},
  \href{http://adsabs.harvard.edu/abs/1997LNP...483..192K}{192--201}

\bibitem[{Klein \& Kerdraon(2011)}]{KleinKerdraon2011}
Klein, K.-L. \& Kerdraon, A. 2011, in 2011 XXXth URSI General Assembly and
  Scientific Symposium\href{https://doi.org/10.1109/ursigass.2011.6051215}{ ({
  IEEE})}

\bibitem[{{Kontogeorgos} {et~al.}(2006{\natexlab{a}}){Kontogeorgos},
  {Tsitsipis}, {Caroubalos}, {Moussas}, {Preka-Papadema}, {Hilaris}, \&
  \etal}]{Kontogeorgos06}
{Kontogeorgos}, A., {Tsitsipis}, P., {Caroubalos}, C., {et~al.}
  2006{\natexlab{a}},
  \href{http://dx.doi.org/10.1007/s10686-006-9066-x}{\color{magenta}Experimental
  Astronomy}, \href{http://adsabs.harvard.edu/abs/2006ExA....21...41K}{21, 41}

\bibitem[{{Kontogeorgos} {et~al.}(2006{\natexlab{b}}){Kontogeorgos},
  {Tsitsipis}, {Moussas}, {Preka-Papadema}, {Hillaris}, {Caroubalos}, \&
  \etal}]{Kontogeorgos}
{Kontogeorgos}, A., {Tsitsipis}, P., {Moussas}, X., {et~al.}
  2006{\natexlab{b}},
  \href{http://dx.doi.org/10.1007/s11214-006-7492-8}{\color{magenta}\ssr},
  \href{http://adsabs.harvard.edu/abs/2006SSRv..122..169K}{122, 169}

\bibitem[{{Krueger}(1979)}]{Krueger79}
{Krueger}, A. 1979, {Introduction to solar radio astronomy and radio physics}
  (Geophysics and Astrophysics Monographs, Dordrecht: Reidel, 1979)

\bibitem[{Magdaleni{\'{c}} {et~al.}(2020)Magdaleni{\'{c}}, Marqu{\'{e}},
  Fallows, Mann, Vocks, Zucca, Dabrowski, Krankowski, \&
  Melnik}]{Magdalenic2020}
Magdaleni{\'{c}}, J., Marqu{\'{e}}, C., Fallows, R.~A., {et~al.} 2020,
  \href{http://dx.doi.org/10.3847/2041-8213/ab9abc}{\color{magenta}\apj}, 897,
  897

\bibitem[{{Pick} \& {Vilmer}(2008)}]{Pick08}
{Pick}, M. \& {Vilmer}, N. 2008,
  \href{http://dx.doi.org/10.1007/s00159-008-0013-x}{\color{magenta}\aapr},
  \href{http://adsabs.harvard.edu/abs/2008A%26ARv..16....1P}{16, 1}

\bibitem[{{Roberts}(1959)}]{Roberts59}
{Roberts}, J.~A. 1959,
  \href{http://dx.doi.org/10.1071/PH590327}{\color{magenta}Australian Journal
  of Physics}, \href{http://adsabs.harvard.edu/abs/1959AuJPh..12..327R}{12,
  327}

\bibitem[{{Tan} {et~al.}(2019){Tan}, {Chen}, {Yang}, {Tan}, {Masuda}, {Chen},
  \& {Misawa}}]{TanBaolin2019}
{Tan}, B., {Chen}, N.-h., {Yang}, Y.-h., {et~al.} 2019,
  \href{http://dx.doi.org/10.3847/1538-4357/ab4718}{\color{magenta}\apj}, 885,
  885

\bibitem[{{Veronig} {et~al.}(2006){Veronig}, {Karlick{\'y}}, {Vr{\v s}nak},
  {Temmer}, {Magdaleni{\'c}}, {Dennis}, {Otruba}, \& {P{\"o}tzi}}]{Veronig2006}
{Veronig}, A.~M., {Karlick{\'y}}, M., {Vr{\v s}nak}, B., {et~al.} 2006,
  \href{http://dx.doi.org/10.1051/0004-6361:20053112}{\color{magenta}\aap},
  \href{http://adsabs.harvard.edu/abs/2006A%26A...446..675V}{446, 675}

\bibitem[{{Veronig} {et~al.}(2005){Veronig}, {Vr{\v s}nak}, {Karlick{\'y}},
  {Temmer}, {Magdaleni{\'c}}, {Dennis}, {Otruba}, \& {P{\"o}tzi}}]{Veronig2005}
{Veronig}, A.~M., {Vr{\v s}nak}, B., {Karlick{\'y}}, M., {et~al.} 2005,
  \href{http://adsabs.harvard.edu/abs/2005ESASP.600E..32V}{in ESA Special
  Publication, Vol. 600, The Dynamic Sun: Challenges for Theory and
  Observations}

\bibitem[{{Veselovsky} {et~al.}(2004){Veselovsky}, {Panasyuk}, {Avdyushin},
  {Bazilevskaya}, {Belov}, {Bogachev}, {Bogod}, {Bogomolov}, {Bothmer},
  {Boyarchuk}, {Vashenyuk}, {Vlasov}, {Gnezdilov}, {Gorgutsa}, {Grechnev},
  {Denisov}, {Dmitriev}, {Dryer}, {Yermolaev}, {Eroshenko}, {Zherebtsov},
  {Zhitnik}, {Zhukov}, {Zastenker}, {Zelenyi}, {Zeldovich}, {Ivanov-Kholodnyi},
  {Ignat'ev}, {Ishkov}, {Kolomiytsev}, {Krasheninnikov}, {Kudela}, {Kuzhevsky},
  {Kuzin}, {Kuznetsov}, {Kuznetsov}, {Kurt}, {Lazutin}, {Leshchenko}, {Litvak},
  {Logachev}, {Lawrence}, {Markeev}, {Makhmutov}, {Mitrofanov}, {Mitrofanov},
  {Morozov}, {Myagkova}, {Nusinov}, {Oparin}, {Panasenco}, {Pertsov},
  {Petrukovich}, {Podorol'sky}, {Romashets}, {Svertilov}, {Svidsky},
  {Svirzhevskaya}, {Svirzhevsky}, {Slemzin}, {Smith}, {Sobel'man}, {Sobolev},
  {Stozhkov}, {Suvorova}, {Sukhodrev}, {Tindo}, {Tokhchukova}, {Fomichev},
  {Chashey}, {Chertok}, {Shishov}, {Yushkov}, {Yakovchouk}, \&
  {Yanke}}]{Veselovsky2004}
{Veselovsky}, I.~S., {Panasyuk}, M.~I., {Avdyushin}, S.~I., {et~al.} 2004,
  \href{http://dx.doi.org/10.1023/B:COSM.0000046229.24716.02}{\color{magenta}Cosmic
  Research}, \href{http://adsabs.harvard.edu/abs/2004CosRe..42..435V}{42, 435}

\bibitem[{{Vr{\v s}nak} \& {Cliver}(2008)}]{Vrsnak08}
{Vr{\v s}nak}, B. \& {Cliver}, E.~W. 2008,
  \href{http://dx.doi.org/10.1007/s11207-008-9241-5}{\color{magenta}\solphys},
  \href{http://adsabs.harvard.edu/abs/2008SoPh..tmp..142V}{253, 215}

\bibitem[{{Vr{\v s}nak} {et~al.}(2006){Vr{\v s}nak}, {Warmuth}, {Temmer},
  {Veronig}, {Magdaleni{\'c}}, {Hillaris}, \& {Karlick{\'y}}}]{Vrsnak06}
{Vr{\v s}nak}, B., {Warmuth}, A., {Temmer}, M., {et~al.} 2006,
  \href{http://dx.doi.org/10.1051/0004-6361:20053740}{\color{magenta}\aap},
  \href{http://adsabs.harvard.edu/abs/2006A%26A...448..739V}{448, 739}

\end{thebibliography}

\end{document}